\documentclass[aps,prl,twocolumn,superscriptaddress,notitlepage]{revtex4-1}
\usepackage{graphicx} % support the \includegraphics command and options
\usepackage{epsfig}
\usepackage{epstopdf}
\usepackage{amsmath}
\usepackage{amsfonts}
\usepackage{amssymb} 
\usepackage{color}
\usepackage{multirow}
\begin{document}

%\vspace*{-3cm}\hspace{4.5cm} IRFU-??-??

%\title{New Perspectives for Deeply Virtual Compton Scattering}
%\title{Beam Energy Dependence of Deeply Virtual Compton Scattering on the Proton}
\title{A Glimpse of Gluons through Deeply Virtual Compton Scattering on the Proton}
\author{M.~Defurne}
\email[Corresponding author: ]{maxime.defurne@cea.fr}
\affiliation{Irfu, CEA, Universit\'{e} Paris-Saclay, 91191 Gif-sur-Yvette, France}
\author{A.~Mart\'i Jim\'enez-Arg\"uello}
\affiliation{Institut de Physique Nucl\'eaire CNRS-IN2P3, Orsay, France}
\affiliation{Facultad de F\'isica, Universidad de Valencia, Valencia, Spain}
\author{Z.~Ahmed}
\affiliation{Syracuse University, Syracuse, New York 13244, USA}
\author{H.~Albataineh}
\affiliation{Texas A\&M University-Kingsville, Kingsville, Texas 78363, USA}
\author{K.~Allada}
\affiliation{Massachusetts Institute of Technology,Cambridge, Massachusetts 02139, USA}
\author{K.~A.~Aniol}
\affiliation{California State University, Los Angeles, Los Angeles, California 90032, USA}
\author{V.~Bellini}
\affiliation{INFN/Sezione di Catania, 95125 Catania, Italy}
\author{M.~Benali}
\affiliation{Clermont universit\'{e}, universit\'{e} Blaise Pascal, CNRS/IN2P3, Laboratoire de physique corpusculaire, FR-63000 Clermont-Ferrand, France}
\author{W.~Boeglin}
\affiliation{Florida International University, Miami, Florida 33199, USA}
\author{P.~Bertin}
\affiliation{Clermont universit\'{e}, universit\'{e} Blaise Pascal, CNRS/IN2P3, Laboratoire de physique corpusculaire, FR-63000 Clermont-Ferrand, France}
\affiliation{Thomas Jefferson National Accelerator Facility, Newport News, Virginia 23606, USA}
\author{M.~Brossard}
\affiliation{Clermont universit\'{e}, universit\'{e} Blaise Pascal, CNRS/IN2P3, Laboratoire de physique corpusculaire, FR-63000 Clermont-Ferrand, France}
\author{A.~Camsonne}
\affiliation{Thomas Jefferson National Accelerator Facility, Newport News, Virginia 23606, USA}
\author{M.~Canan}
\affiliation{Old Dominion University, Norfolk, Virginia 23529, USA}
\author{S.~Chandavar}
\affiliation{Ohio University, Athens, Ohio 45701, USA}
\author{C.~Chen}
\affiliation{Hampton University, Hampton, Virginia 23668, USA}
\author{J.-P.~Chen}
\affiliation{Thomas Jefferson National Accelerator Facility, Newport News, Virginia 23606, USA}
\author{C.W.~de~Jager}
\thanks{Deceased}
\affiliation{Thomas Jefferson National Accelerator Facility, Newport News, Virginia 23606, USA}
\author{R.~de~Leo}
\affiliation{Universit\`{a} di Bari, 70121 Bari, Italy}
\author{C.~Desnault}
\affiliation{Institut de Physique Nucl\'eaire CNRS-IN2P3, Orsay, France}
\author{A.~Deur}
\affiliation{Thomas Jefferson National Accelerator Facility, Newport News, Virginia 23606, USA}
\author{L.~El~Fassi}
\affiliation{Rutgers, The State University of New Jersey, Piscataway, New Jersey 08854, USA}
\author{R.~Ent}
\affiliation{Thomas Jefferson National Accelerator Facility, Newport News, Virginia 23606, USA}
\author{D.~Flay}
\affiliation{Temple University, Philadelphia, Pennsylvania 19122, USA}
\author{M.~Friend}
\affiliation{Carnegie Mellon University, Pittsburgh, Pennsylvania 15213, USA}
\author{E.~Fuchey}
\affiliation{Irfu, CEA, Universit\'{e} Paris-Saclay, 91191 Gif-sur-Yvette, France}
\affiliation{Clermont universit\'{e}, universit\'{e} Blaise Pascal, CNRS/IN2P3, Laboratoire de physique corpusculaire, FR-63000 Clermont-Ferrand, France}
\affiliation{University of Connecticut, Storrs, Connecticut 06269, USA}
\author{S.~Frullani}
\thanks{Deceased}
\affiliation{INFN/Sezione Sanit\`{a}, 00161 Roma, Italy}
\author{F.~Garibaldi}
\affiliation{INFN/Sezione Sanit\`{a}, 00161 Roma, Italy}
\author{D.~Gaskell}
\affiliation{Thomas Jefferson National Accelerator Facility, Newport News, Virginia 23606, USA}
\author{A.~Giusa}
\affiliation{INFN/Sezione di Catania, 95125 Catania, Italy}
\author{O.~Glamazdin}
\affiliation{Kharkov Institute of Physics and Technology, Kharkov 61108, Ukraine}
\author{S.~Golge}
\affiliation{North Carolina Central University, Durham, North Carolina 27701, USA}
\author{J.~Gomez}
\affiliation{Thomas Jefferson National Accelerator Facility, Newport News, Virginia 23606, USA}
\author{O.~Hansen}
\affiliation{Thomas Jefferson National Accelerator Facility, Newport News, Virginia 23606, USA}
\author{D.~Higinbotham}
\affiliation{Thomas Jefferson National Accelerator Facility, Newport News, Virginia 23606, USA}
\author{T.~Holmstrom}
\affiliation{Longwood University, Farmville, Virginia 23909, USA}
\author{T.~Horn}
\affiliation{The Catholic University of America, Washington, DC 20064, USA}
\author{J.~Huang}
\affiliation{Massachusetts Institute of Technology,Cambridge, Massachusetts 02139, USA}
\author{M.~Huang}
\affiliation{Duke University, Durham, North Carolina 27708, USA}
\author{C.E.~Hyde}
\affiliation{Old Dominion University, Norfolk, Virginia 23529, USA}
\affiliation{Clermont universit\'{e}, universit\'{e} Blaise Pascal, CNRS/IN2P3, Laboratoire de physique corpusculaire, FR-63000 Clermont-Ferrand, France}
\author{S.~Iqbal}
\affiliation{California State University, Los Angeles, Los Angeles, California 90032, USA}
\author{F.~Itard}
\affiliation{Clermont universit\'{e}, universit\'{e} Blaise Pascal, CNRS/IN2P3, Laboratoire de physique corpusculaire, FR-63000 Clermont-Ferrand, France}
\author{H.~Kang}
\affiliation{Seoul National University, Seoul, South Korea}
\author{A.~Kelleher}
\affiliation{College of William and Mary, Williamsburg, Virginia 23187, USA}
\author{C.~Keppel}
\affiliation{Thomas Jefferson National Accelerator Facility, Newport News, Virginia 23606, USA}
\author{S.~Koirala}
\affiliation{Old Dominion University, Norfolk, Virginia 23529, USA}
\author{I.~Korover}
\affiliation{Tel Aviv University, Tel Aviv 69978, Israel}
\author{J.J.~LeRose}
\affiliation{Thomas Jefferson National Accelerator Facility, Newport News, Virginia 23606, USA}
\author{R.~Lindgren}
\affiliation{University of Virginia, Charlottesville, Virginia 22904, USA}
\author{E.~Long}
\affiliation{Kent State University, Kent, Ohio 44242, USA}
\author{M.~Magne}
\affiliation{Clermont universit\'{e}, universit\'{e} Blaise Pascal, CNRS/IN2P3, Laboratoire de physique corpusculaire, FR-63000 Clermont-Ferrand, France}
\author{J.~Mammei}
\affiliation{University of Massachusetts, Amherst, Massachusetts 01003, USA}
\author{D.J.~Margaziotis}
\affiliation{California State University, Los Angeles, Los Angeles, California 90032, USA}
\author{P.~Markowitz}
\affiliation{Florida International University, Miami, Florida 33199, USA}
\author{M.~Mazouz}
\affiliation{Facult\'e des Sciences de Monastir, 5000 Tunisia}
\author{F.~Meddi}
\affiliation{INFN/Sezione Sanit\`{a}, 00161 Roma, Italy}
\author{D.~Meekins}
\affiliation{Thomas Jefferson National Accelerator Facility, Newport News, Virginia 23606, USA}
\author{R.~Michaels}
\affiliation{Thomas Jefferson National Accelerator Facility, Newport News, Virginia 23606, USA}
\author{M.~Mihovilovic}
\affiliation{University of Ljubljana, 1000 Ljubljana, Slovenia}
\author{C.~Mu\~noz~Camacho}
\affiliation{Clermont universit\'{e}, universit\'{e} Blaise Pascal, CNRS/IN2P3, Laboratoire de physique corpusculaire, FR-63000 Clermont-Ferrand, France}
\affiliation{Institut de Physique Nucl\'eaire CNRS-IN2P3, Orsay, France}
\author{P.~Nadel-Turonski}
\affiliation{Thomas Jefferson National Accelerator Facility, Newport News, Virginia 23606, USA}
\author{N.~Nuruzzaman}
\affiliation{Hampton University, Hampton, Virginia 23668, USA}
\author{R.~Paremuzyan}
\affiliation{Institut de Physique Nucl\'eaire CNRS-IN2P3, Orsay, France}
\author{A.~Puckett}
\affiliation{Los Alamos National Laboratory, Los Alamos, New Mexico 87545, USA}
\author{V.~Punjabi}
\affiliation{Norfolk State University, Norfolk, Virginia 23529, USA}
\author{Y.~Qiang}
\affiliation{Thomas Jefferson National Accelerator Facility, Newport News, Virginia 23606, USA}
\author{A.~Rakhman}
\affiliation{Syracuse University, Syracuse, New York 13244, USA}
\author{M.N.H.~Rashad}
\affiliation{Old Dominion University, Norfolk, Virginia 23529, USA}
\author{S.~Riordan}
\affiliation{Stony Brook University, Stony Brook, New York 11794, USA}
\author{J.~Roche}
\affiliation{Ohio University, Athens, Ohio 45701, USA}
\author{G.~Russo}
\affiliation{INFN/Sezione di Catania, 95125 Catania, Italy}
\author{F.~Sabati\'e}
\affiliation{Irfu, CEA, Universit\'{e} Paris-Saclay, 91191 Gif-sur-Yvette, France}
\author{K.~Saenboonruang}
\affiliation{University of Virginia, Charlottesville, Virginia 22904, USA}
\affiliation{Kasetsart University, Chatuchak, Bangkok, 10900, Thailand}
\author{A.~Saha}
\thanks{Deceased}
\affiliation{Thomas Jefferson National Accelerator Facility, Newport News, Virginia 23606, USA}
\author{B.~Sawatzky}
\affiliation{Thomas Jefferson National Accelerator Facility, Newport News, Virginia 23606, USA}
\affiliation{Temple University, Philadelphia, Pennsylvania 19122, USA}
\author{L.~Selvy}
\affiliation{Kent State University, Kent, Ohio 44242, USA}
\author{A.~Shahinyan}
\affiliation{Yerevan Physics Institute, Yerevan 375036, Armenia}
\author{S.~Sirca}
\affiliation{University of Ljubljana, 1000 Ljubljana, Slovenia}
\author{P.~Solvignon}
\thanks{Deceased}
\affiliation{Thomas Jefferson National Accelerator Facility, Newport News, Virginia 23606, USA}
\affiliation{University of New Hampshire, Durham, New Hampshire 03824, USA}
\author{M.L.~Sperduto}
\affiliation{INFN/Sezione di Catania, 95125 Catania, Italy}
\author{R.~Subedi}
\affiliation{George Washington University, Washington, DC 20052, USA}
\author{V.~Sulkosky}
\affiliation{Massachusetts Institute of Technology,Cambridge, Massachusetts 02139, USA}
\author{C.~Sutera}
\affiliation{INFN/Sezione di Catania, 95125 Catania, Italy}
\author{W.A.~Tobias}
\affiliation{University of Virginia, Charlottesville, Virginia 22904, USA}
\author{G.M.~Urciuoli}
\affiliation{INFN/Sezione di Roma, 00185 Roma, Italy}
\author{D.~Wang}
\affiliation{University of Virginia, Charlottesville, Virginia 22904, USA}
\author{B.~Wojtsekhowski}
\affiliation{Thomas Jefferson National Accelerator Facility, Newport News, Virginia 23606, USA}
\author{H.~Yao}
\affiliation{Temple University, Philadelphia, Pennsylvania 19122, USA}
\author{Z.~Ye}
\affiliation{University of Virginia, Charlottesville, Virginia 22904, USA}
\author{X.~Zhan}
\affiliation{Argonne National Laboratory, Lemont, Illinois 60439, USA}
\author{J.~Zhang}
\affiliation{Thomas Jefferson National Accelerator Facility, Newport News, Virginia 23606, USA}
\author{B.~Zhao}
\affiliation{College of William and Mary, Williamsburg, Virginia 23187, USA}
\author{Z.~Zhao}
\affiliation{University of Virginia, Charlottesville, Virginia 22904, USA}
\author{X.~Zheng}
\affiliation{University of Virginia, Charlottesville, Virginia 22904, USA}
\author{P.~Zhu}
\affiliation{University of Virginia, Charlottesville, Virginia 22904, USA}
\collaboration{The Jefferson Lab Hall A Collaboration}

\date{\today}

\begin{abstract}
 The proton is composed of quarks and gluons, bound by the most elusive mechanism of strong interaction called confinement. In this work, the dynamics of quarks and gluons are investigated using deeply virtual Compton scattering (DVCS): produced by a multi-GeV electron, a highly virtual photon scatters off the proton which subsequently radiates a high energy photon. Similarly to holography, measuring not only the magnitude but also the phase of the DVCS amplitude allows to perform 3D images of the internal structure of the proton. The phase is made accessible through the quantum-mechanical interference of DVCS with the Bethe-Heitler (BH) process, in which the final photon is emitted by the electron rather than the proton. 

We report herein the first full determination of the BH-DVCS interference by exploiting the distinct energy dependences of the DVCS and BH amplitudes. In the high energy regime where the scattering process is expected to occur off a single quark in the proton, these accurate measurements show an intriguing sensitivity to gluons, the carriers of the strong interaction. 
\end{abstract}


\maketitle

%%%%%%%%%%%%%%%%%
The dynamics of quarks and gluons inside the nucleon are governed by the strong interaction, described by the theory of Quantum Chromodynamics (QCD). At a distance scale of the nucleon radius, perturbative computations cannot be performed because of the large value of the strong coupling constant $\alpha_S$. To unravel the internal dynamics of the nucleon and answer fundamental questions from the origin of its spin to the mechanism of confinement, lepton scattering experiments have proven to be a powerful tool. Indeed, elastic scattering allows to access the transverse spatial distribution of charge and current in the nucleon through measurements of its electric and magnetic form factors, whereas parton distribution functions measured in deep inelastic experiments provide information on the longitudinal momentum carried by the confined quarks and gluons. Developed in the mid-90s, the Generalized Parton Distributions (GPDs)~\cite{Mueller:1998fv, Ji:1996ek, Radyushkin:1997ki} provide a higher level of information and encode correlations between the transverse position and the longitudinal momentum of quarks and gluons inside the nucleon~\cite{Ji:1996nm}. Being a $\frac{1}{2}$-spin particle, the nucleon is described by 4 chiral-even GPDs $\{H,\,E,\,\widetilde{H},\,\widetilde{E}\}$ and their chiral-odd counterparts more commonly called ``transversity GPDs'', 
for each quark flavor and for gluons.  

\begin{figure}[bp]
\begin{center}
\includegraphics[width=\linewidth]{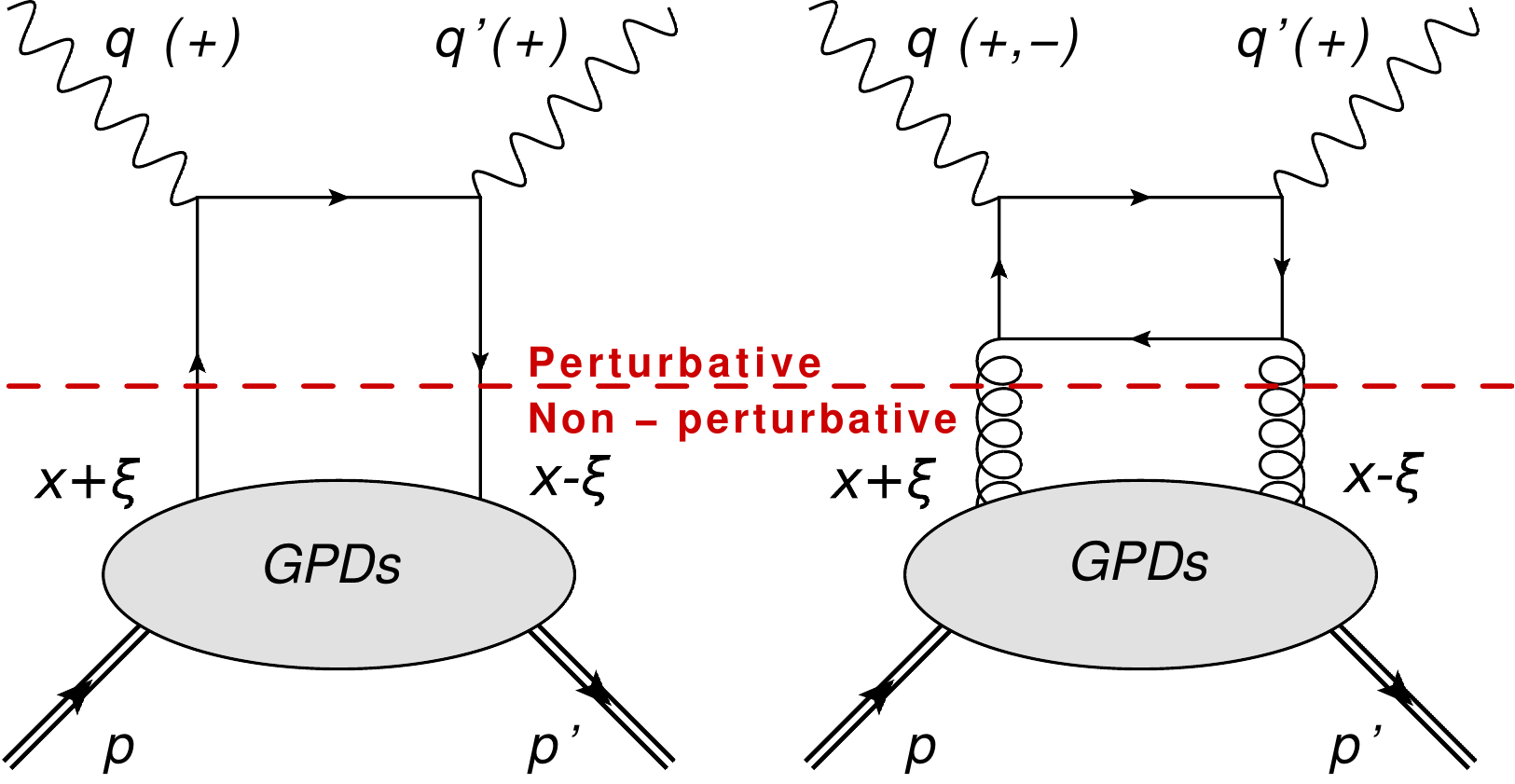}
\caption{\textbf{Leading-twist DVCS diagrams.} At leading-order in perturbative QCD (left), the virtual photon with four-momentum $q$ interacts with a single quark (single straight line) from the proton $p$, in the limit $Q^2=-q^2$ much larger than the proton mass squared. Subsequently, the active quark emits a real photon with four-momentum $q'$. The recoil proton has four-momentum $p'$. Perturbation theory can be used to calculate the part of the amplitude above the (dashed) factorization line, whereas GPDs encode the non-perturbative structure of the nucleon. At next-to-leading order in perturbative QCD (right), a gluon (curly line) from the proton splits into a quark-antiquark pair and the quark absorbs the virtual photon. The average longitudinal momentum fraction carried by the active parton (quark/gluon) is $x$ and -2$\xi$ is the longitudinal momentum transfer. The helicity of the photons contributing to the leading-twist amplitudes are specified in parenthesis.}
\label{fig:LO-NLO}
\end{center}
\end{figure}

GPDs are accessible through deep virtual photoproduction processes: a high-energy virtual photon scatters off the proton and all the subsequent particles in the final state are identified~\cite{Aktas:2005ty,Adloff:2001cn,Chekanov:2003ya, Airapetian:2008aa,Collaboration:2010kna,Mazouz:2007aa,Defurne:2015kxq,Jo:2015ema}. The high-energy scale introduced by the virtual photon (or equivalently its short distance resolution scale) ensures that the reaction is governed by perturbative dynamics of quarks and gluons ($\alpha_S\ll 1$). In this work we focus on the process in which a single high-energy photon is emitted by the scattered proton, the so-called deeply virtual Compton scattering (DVCS), described in Fig.~\ref{fig:LO-NLO}.

Collinear factorization theorems~\cite{Collins:1996fb,Ji:1998xh} demonstrate that at sufficiently high energy, the DVCS amplitude  is a convolution of a perturbative kernel with 
the GPDs of the nucleon --- which describe the  nucleon's non-perturbative structure  (see Fig.~\ref{fig:LO-NLO}). 
These convolutions, called Compton Form Factors (CFFs), can be classified according to photon helicity states. With $\lambda$ and $\lambda'$  the helicity state of the virtual photon and outgoing real photon, respectively,  we distinguish 3 kinds of photon-helicity-dependent CFFs $\mathcal{F}_{\lambda \lambda'}\in \left\{\mathcal{H}_{\lambda \lambda'},\mathcal{E}_{\lambda \lambda'},\widetilde{\mathcal{H}}_{\lambda \lambda'}, \widetilde{\mathcal{E}}_{\lambda \lambda'}\right\}$~\cite{Belitsky:2012ch}:
\begin{itemize}
\item The helicity-conserved CFFs ($\mathcal{F}_{++}$) which describe diagrams for which the virtual and the outgoing photons have the same helicity state.
\item The longitudinal-to-transverse helicity-flip CFFs ($\mathcal{F}_{0+}$) describing the contribution of a longitudinally polarized virtual photon. 
\item The transverse-to-transverse helicity-flip CFFs ($\mathcal{F}_{-+}$) for which the virtual and the outgoing photons have opposite helicities.
\end{itemize}
The CFFs are also classified according to the inverse power of $Q\equiv \sqrt{Q^2}$ with which they enter the DVCS amplitude.
This power is called the twist, and is equal to the dimension minus the spin of the corresponding operator. The leading twist CFFs are $\mathcal{F}_{++}$ and $\mathcal{F}_{-+}$, which are twist-2. CFFs $\mathcal{F}_{0+}$ are twist-3, \emph{i.e.} $\frac{1}{Q}$-suppressed with respect to the leading-twist CFFs. Note that 
the gluon contribution (Fig.~\ref{fig:LO-NLO}) while twist-2, is suppressed by a factor of $\alpha_S$ (Next-to-Leading-Order).
%$\alpha_S$ which prevents any gluon contribution, $\mathcal{F}_{-+}$ are twist-4 and consequently $\frac{1}{Q^2}$-suppressed with respect to the leading-twist CFFs.  

\begin{figure}[bt]
\begin{center}
\includegraphics[width=\linewidth]{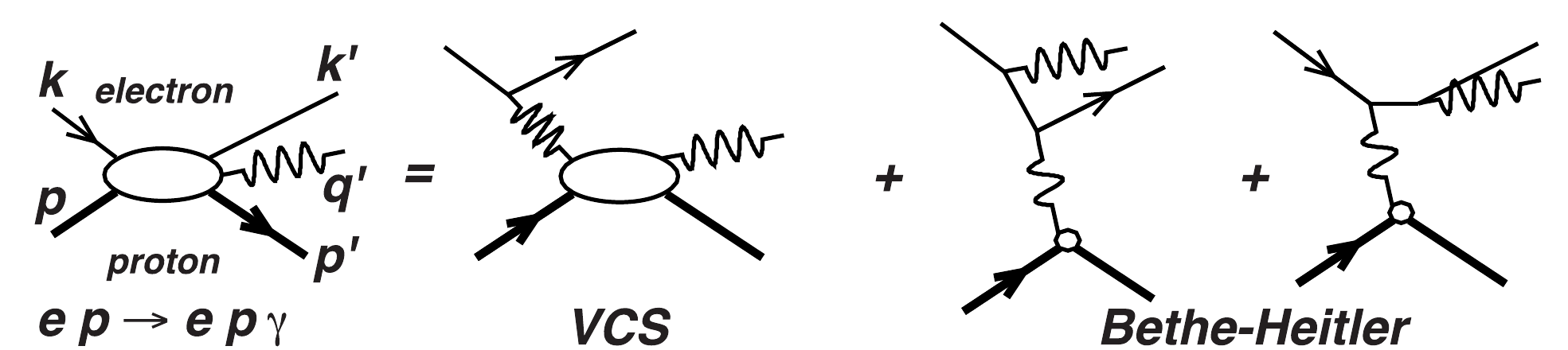}
\caption{
\textbf{Lowest-order Quantum Electrodynamics amplitude for the $\boldsymbol{ep\rightarrow ep\gamma}$ reaction.}
 The momentum four-vectors of external particles are labeled on the left.
The net four-momentum transfer to the proton is 
$\Delta_\mu=(q-q')_\mu=(p'-p)_\mu$. In the virtual Compton scattering
(VCS) amplitude, the (spacelike) virtuality of the incident photon is
$Q^2=-q^2=-(k-k')^2$.  The Bjorken variable $x_B$ is defined as $x_B = Q^2/(2q\cdot P)$. In the Bethe-Heitler amplitude, the virtuality
of the incident photon is $-\Delta^2=-t$.}
\label{fig:epepg}
\end{center}
\end{figure}

To experimentally study DVCS, the virtual photon in the initial state is produced via the scattering of a multi-GeV electron off a proton. Consequently, DVCS events have an electron and a proton ($ep$) in the initial state, with a final state composed of the scattered electron, the recoil proton and the high energy photon ($ep\gamma$). However, the final photon of the reaction $ep\rightarrow ep\gamma$ can also be emitted by either the incoming or scattered electron instead of the proton, a competing channel called Bethe-Heitler (BH). Therefore, the exclusive photon electroproduction $ep\rightarrow ep\gamma$ cross section of a polarized electron beam of energy $k$  off an unpolarized target of mass $M$ (see Fig. \ref{fig:epepg}) can be written as~\cite{Belitsky:2010jw}: 
\begin{multline}
\frac{d^4\sigma(h)}{dQ^2 dx_B dt d\phi} =
  \frac{d^2\sigma_0}{dQ^2 dx_B} \times\\  \left[
       \left|\mathcal T^{BH} \right|^2 +
        \left| \mathcal T^{DVCS}(h) \right|^2 -
        \mathcal I(h)\right] \label{eq:dsigDVCS}
\end{multline}
%\begin{eqnarray}
% \frac{d^2\sigma_0}{dQ^2 dx_B} &=&
%\frac{\alpha_{\rm QED}^3}{8\pi(s_e-M^2)^2 x_B} 
%\frac{1}{\sqrt{1+\epsilon^2}}   \\
%\epsilon^2 &=& 4 M^2 x_B^2/Q^2 \nonumber% \\
%s_e&=& 2 M k + M^2 \nonumber 
%\label{eq:dsig0}
%\end{eqnarray}
where $\phi$ is the angle between the leptonic and hadronic planes defined by the Trento convention~\cite{Bacchetta:2004jz}, $h$ the lepton helicity and  $\mathcal I$ is the interference of the virtual Compton $\mathcal T^{DVCS}$ and Bethe-Heitler $\mathcal T^{BH}$ amplitudes.
 %%%%%%%%%%%%%%%%%%%

The interference between BH and DVCS provides a way to independently access the real and imaginary parts of CFFs. At leading-order, the imaginary part of $\mathcal{F}_{++}$ is directly related to the corresponding GPD at $x=\xi$:
\begin{eqnarray}
\mathcal R\text{e}\,\mathcal{F}_{++}&=& \mathcal{P} \int_{-1}^{1}dx\left[\frac{1}{x-\xi} -\kappa \frac{1}{x+\xi}\right]F(x,\xi,t)\;,\nonumber\\
\mathcal I\text{m}\,\mathcal{F}_{++}&=& -\pi\left[F(\xi,\xi,t)+\kappa F(-\xi,\xi,t)\right]\;,
\label{eq:dispersion}
\end{eqnarray}
where $\kappa=-1$ if $F \in\{H,E\}$ and $1$ if $F \in\{\widetilde{H},\widetilde{E}\}$. Recent phenomenology uses the leading-twist (LT) and leading-order (LO) approximation in order to extract or parametrize GPDs, which translates into neglecting $\mathcal{F}_{0+}$ and $\mathcal{F}_{-+}$ and using the relations of Eq.~\ref{eq:dispersion}~\cite{Kumericki:2009uq, Kumericki:2016ehc, Dupre:2016mai}.

We report herein measurements of helicity-dependent and helicity-independent photon electroproduction cross sections with high statistical accuracy in Hall A of Jefferson Lab.   %The specific kinematics are summarized in Tab.~\ref{tbl:DVCSKin}.   
The H$(\vec{e},e'\gamma)p$ cross section was measured at six kinematic points centered at $x_B=0.36$.  At $Q^2=1.5\text{ GeV}^2$, the incident beam energies
were 3.355 and 5.550 GeV, and we report cross sections in three bins of $-t$, centered at $0.18,\, 0.24,$ and $0.30\text{ GeV}^2$.  
At $Q^2 = 1.75$ and $2.0\text{ GeV}^2$, the incident beam energies were 4.455 and 5.550 GeV, and we report four bins in $-t$, centered at $0.18,\, 0.24,$ and $0.30$ and $0.36\text{ GeV}^2$.
The present experimental study was initiated to separate the DVCS-BH interference and DVCS$^2$ contributions to the $ep\to ep\gamma$ cross section, by exploiting the different energy dependences of  the BH and DVCS amplitudes.  %In particular, 
%the real part of this interference has only previously been accessed by measuring the  asymmetry between incident electron and positron beams
Until now, only the asymmetry between incident electron and positron beams could be used to constrain the real part of this interference~\cite{Airapetian:2008aa,Aaron:2009ac}. 

In experiment E07-007 a longitudinally polarized electron beam impinged on a 15-cm-long liquid H$_2$ target. Beam polarization was continuously measured by the Hall A Compton polarimeter and found to be 72$\pm 2\%_\text{sys}$ on average. Scattered electrons were detected in the left High Resolution Spectrometer (HRS)~\cite{Alcorn:2004sb}. Events were triggered by the coincidence of a scintillator plane (S2m) and a signal in a gas \u{C}erenkov counter. The HRS $\delta p/p\sim 10^{-4}$ momentum resolution and $\delta\theta\sim 0.6$~mr horizontal angular resolution provide a precise measurement of the electron kinematics and interaction vertex. Tracking efficiency was known to 0.5\%. The final state photon was detected in an electromagnetic calorimeter consisting on an 16$\times$13 array of PbF$_2$ crystals. Its energy resolution was measured to be 2.4\% at 4.2~GeV, with $\sim$3~mm spatial resolution.

\begin{figure}[bt]
\begin{center}
\includegraphics[width=\linewidth]{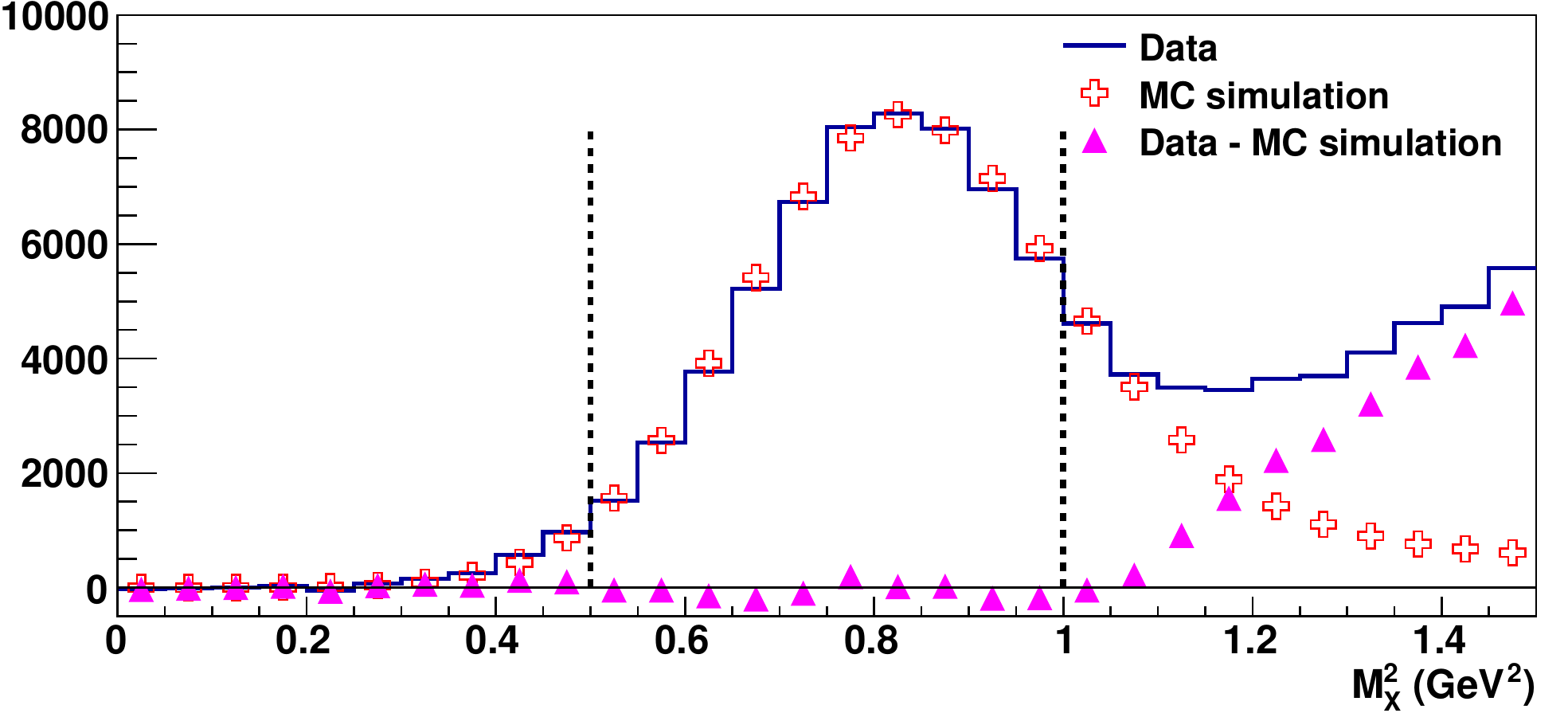}
\caption{\textbf{Missing mass squared distribution at $\boldsymbol{E^{beam}}$=5.55~GeV, $\boldsymbol{Q^2}$=1.75~GeV$\boldsymbol{^2}$} (Color online) The accidentals and the $\pi^0$-contamination have already been subtracted. The Monte-Carlo simulation is represented by the open crosses, whereas the triangles show the estimated inclusive yield obtained by subtracting the simulation from the data. The vertical dotted lines illustrates the two cuts applied on $M^2_X$ in the analysis.}
\label{fig:MX}
\end{center}
\end{figure}

The exclusivity of the reaction is ensured by a cut on the $ep\rightarrow e\gamma X$ missing mass squared $M_X^2=(k+p-k'-q')^2$ (see Fig.~\ref{fig:MX}). The number of events $N_\mathcal{C}$ below the missing mass cut $M^2_\mathcal{C}$ is the sum of four contributions:
\begin{equation}
N_\mathcal{C}=N_{ep\rightarrow ep\gamma}+N_{\pi^0-1\gamma}+N_{acc}+N_{SIDIS}\;,
\end{equation}
with $N_{ep\rightarrow ep\gamma}$ the number of exclusive photon events, $N_{\pi^0-1\gamma}$ the contamination from $\pi^0$ decays that yield only one photon in the calorimeter, $N_{acc}$ the number of electron-photon accidental coincidences and $N_{SIDIS}$ the contamination from semi-inclusive events $ep\to ep\gamma X$. %such as N$_{ep\rightarrow ep\gamma \pi}$,etc. 
The contamination caused by asymmetric $\pi^0$-decays with respect to the pion momentum was estimated by simulating thousands of decays for each $\pi^0$ identified in the data and computing the likelihood for each to yield only one photon in the experimental acceptance. The subtraction of $N_{acc}$ was performed by analyzing events where the scattered electron and the detected photon were not in coincidence. In addition, we applied a 800~MeV energy cut on the photon to remove most of the accidental background and required a value of $M_X^2>0.5$~GeV$^2$ to increase the signal/background ratio~(Fig.~\ref{fig:MX}). We also applied a $M^2_X<$1~GeV$^2$ cut so that $N_{SIDIS}$ is less than 1\% of exclusive $N_{ep\rightarrow ep\gamma}$ events. The significant fraction of exclusive photon events with a missing mass squared higher than $M^2_\mathcal{C}$ is corrected by applying the same cut to the Monte-Carlo simulation used to compute the experimental acceptance. This fraction of events removed varies from bin to bin since the width and position of the exclusive signal may slightly change from one bin to another. The energy resolution of the calorimeter was smeared locally in order to match the missing mass resolution observed in the experimental data, and the systematic uncertainty associated to the exclusivity cut estimated to be 2\%.   

The Monte-Carlo simulation is based on the GEANT4 toolkit and includes real and virtual radiative corrections following the procedure described in~\cite{Defurne:2015kxq} and based on calculations by Vanderhaeghen~\emph{et al.}~\cite{Vanderhaeghen:2000ws}. A 2\%-systematic uncertainty has been attributed to the radiative corrections and a 1\%-uncertainty to the HRS acceptance model~\cite{Rvachev:2001}. The simulation is used to account for bin migration effects due to detector resolution and Bremsstrahlung radiation, with 1\% systematic uncertainty. An additional bin in $t$ is used to correct for bin migration in and out of the largest $|t|-$bin. We also include 2\% uncertainty for the integrated luminosity and data acquisition dead-time correction and 0.5\% for trigger efficiency, which yields a total systematic uncertainty of 3.7\% for the unpolarized cross sections and 4.2\% for the helicity-dependent cross sections.

\begin{figure*}[bth]
\begin{center}
\centering\includegraphics[width=0.9\linewidth]{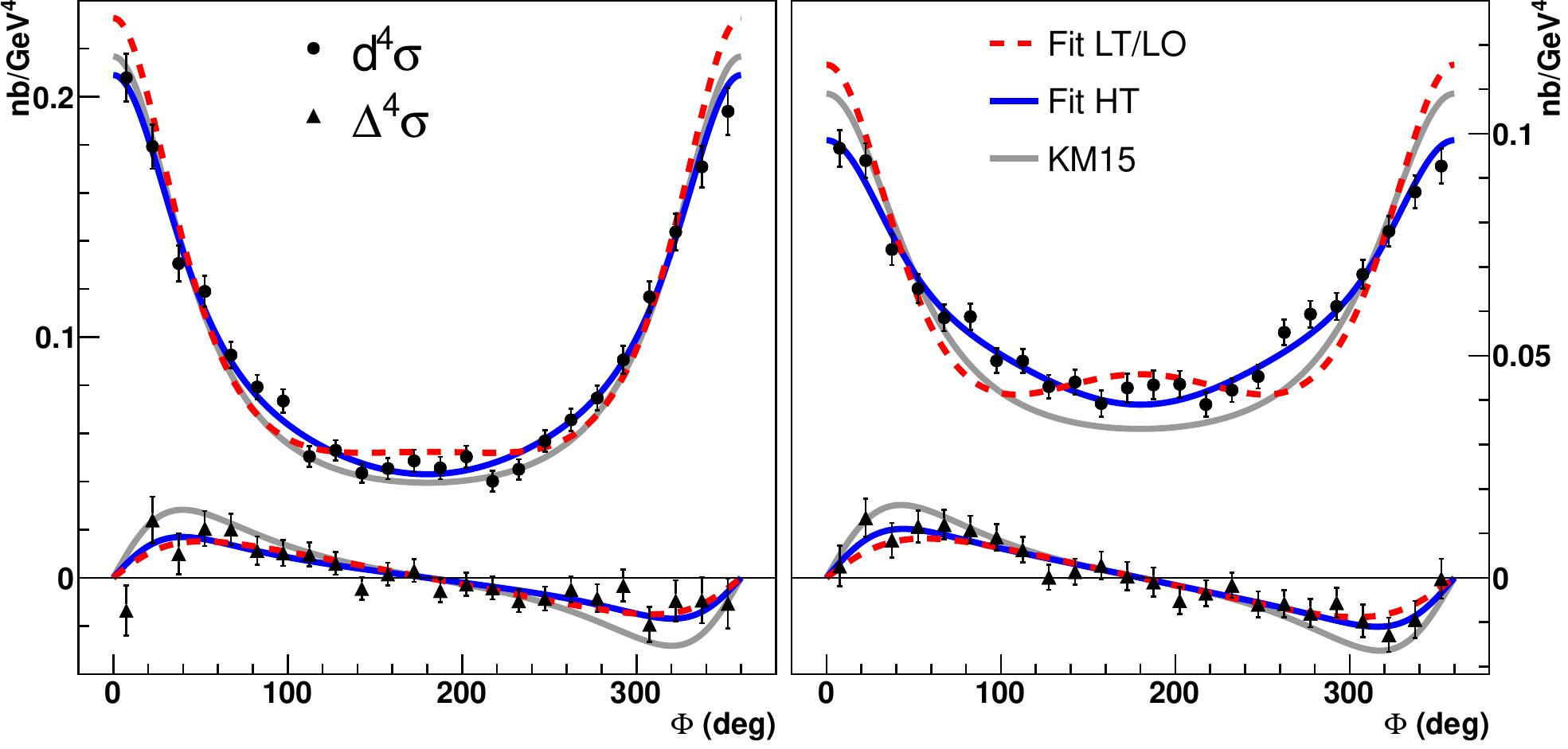}
\caption{\textbf{Beam helicity-dependent ($\boldsymbol{\Delta^4\sigma}$) and helicity-independent ($\boldsymbol{d^4\sigma}$) cross sections at $\boldsymbol{Q^2}$=1.75~GeV$\boldsymbol{^2}$, $\boldsymbol{x_B}$=0.36, and $\boldsymbol{t=-0.30}$~GeV$\boldsymbol{^2}$.} (Color online) The beam energies are $E^{beam}$=4.455~GeV (left) and $E^{beam}$=5.55~GeV (right). Dashed lines represent the result of the LT/LO fit with $\mathbb{H}_{++}$,$\mathbb{E}_{++}$,$\widetilde{\mathbb{H}}_{++}$ and $\widetilde{\mathbb{E}}_{++}$. Solid lines show the result of the HT fit with $\mathbb{H}_{++}$,$\widetilde{\mathbb{H}}_{++}$,$\mathbb{H}_{0+}$, and $\widetilde{\mathbb{H}}_{0+}$. Curves for the NLO fit ($\mathbb{H}_{++}$,$\widetilde{\mathbb{H}}_{++}$,$\mathbb{H}_{-+}$, and $\widetilde{\mathbb{H}}_{-+}$) overlap with the HT fit and are not shown. Results from the KM15~\cite{Kumericki:2009uq, Kumericki:2016ehc} fit to previously published DVCS data is also presented.}
\label{fig:Fit_result}
\end{center}
\end{figure*}

The scattering amplitude is a Lorentz invariant quantity, but the deeply virtual scattering process nonetheless defines a preferred axis (light-cone axis) for describing the scattering process. At finite $Q^2$ and non-zero $t$, there is an ambiguity in defining this axis, though all definitions converge as
$Q^2\to \infty$ at fixed $t$. Belitsky et al.~\cite{Belitsky:2012ch} decompose the DVCS amplitude in terms of photon-helicity states where the light-cone axis is defined in the plane of the four-vectors $q$ and $P$.  This leads to the CFFs defined previously. Recently, Braun \emph{et al.}~\cite{Braun:2014sta} proposed an alternative decomposition which defines the light cone axis in the plane formed by $q$ and $q'$ and argue that this is  more convenient to account for kinematical power corrections of $\mathcal{O}(t/Q^2)$ and $\mathcal{O}(M^2/Q^2)$. The bulk of these corrections can be included by rewriting the CFFs $\mathcal{F}_{\lambda \lambda'}$ in terms of $\mathbb{F}_{\lambda \lambda'}$ using the following map~\cite{Braun:2014sta}:
\begin{eqnarray}
\label{eq:BMPtoBMJ}
\mathcal{F}_{++}=&\mathbb{F}_{++}+\frac{\chi}{2}\left[\mathbb{F}_{++}+ \mathbb{F}_{-+}\right]-\chi_0 \mathbb{F}_{0+}\;,\\
\mathcal{F}_{-+}=&\mathbb{F}_{-+}+\frac{\chi}{2}\left[\mathbb{F}_{++}+ \mathbb{F}_{-+}\right]-\chi_0 \mathbb{F}_{0+}\;,\label{eq:BMPtoBMJ1}\\
\mathcal{F}_{0+}=&-(1+\chi)\mathbb{F}_{0+}+\chi_0\left[\mathbb{F}_{++}+ \mathbb{F}_{-+}\right]\;,\label{eq:BMPtoBMJ2}
\end{eqnarray}
where kinematic parameters $\chi_0$ and $\chi$ are defined  as follows (Eq.~48 of Ref~\cite{Braun:2014sta}):
\begin{eqnarray}
\chi_0=&\displaystyle\frac{\sqrt{2}Q\widetilde{K}}{\sqrt{1+\epsilon^2}(Q^2+t)}&\propto \frac{\sqrt{t_{min}-t}}{Q}\;,\\
\chi=&\displaystyle\frac{Q^2-t+2x_Bt}{\sqrt{1+\epsilon^2}(Q^2+t)}-1&\propto \frac{t_{min}-t}{Q^2}\;.
\end{eqnarray}

Within the $\mathbb{F}_{\mu \nu}$-parameterization, the leading-twist and leading-order approximation consists in keeping $\mathbb{F}_{++}$ and neglecting both $\mathbb{F}_{0+}$ and  $\mathbb{F}_{-+}$. Nevertheless, as a consequence of Eq.~\ref{eq:BMPtoBMJ1} and \ref{eq:BMPtoBMJ2}, $\mathcal{F}_{0+}$ and $\mathcal{F}_{-+}$ are no longer equal to zero since proportional to $\mathbb{F}_{++}$. The functions that can be extracted from data to describe the three dimensional structure of the nucleon become:
\begin{equation}
\mathcal{F}_{++}=(1+\frac{\chi}{2})\mathbb{F}_{++},\,
\mathcal{F}_{0+}=\chi_0\mathbb{F}_{++},\,
\mathcal{F}_{-+}=\frac{\chi}{2}\mathbb{F}_{++}.
\end{equation}
A numerical application gives $\chi_0=$0.25 and $\chi=$0.06 for $Q^2$=2~GeV$^2$, x$_B$=0.36 and $t=-0.24$~GeV$^2$. Considering the large size of the parameters $\chi_0$ and $\chi$, these kinematical power corrections cannot be neglected in precision DVCS phenomenology, in particular in order to separate the DVCS-BH interference and DVCS$^2$ contributions. Indeed, when the beam energy changes, not only do the contributions of the DVCS-BH interference and DVCS$^2$ terms change but also the polarization of the virtual photon changes, 
thereby modifying the weight of the different helicity amplitudes.

Fig.~\ref{fig:Fit_result} shows the beam helicity-dependent and helicity-independent cross sections measured in one kinematic bin, at two different values of the incident beam energy. Neglecting the (logarithmic) $Q^2$--evolution of the CFFs between 1.5 to 2~GeV$^2$, a combined fit of all the data at constant $x_B$ and $t$ is performed. 
For each $-t$ bin, this fit includes the helicity-dependent and helicity-independent cross sections at 2 values of beam energy and all 3 values of $Q^2$.
% (see Tab.~\ref{tbl:DVCSKin}). 
The leading-twist and leading-order (LO/LT) fit is shown in Fig.~\ref{fig:Fit_result} for $t=-0.30$~GeV$^2$, in which the free parameters are the real and imaginary parts of $\mathbb{H}_{++}$, $\widetilde{\mathbb{H}}_{++}$, $\mathbb{E}_{++}$ and $\widetilde{\mathbb{E}}_{++}$. This fit reproduces very poorly the angular distribution of the data yielding a value of $\chi^2/ndf=504/208$. Indeed, the strong enhancement of the $\cos{\phi}$-harmonics in the DVCS$^2$ amplitude originated by the large size of $\chi_0$ translates into the bump in the dashed line around $\phi$=180$^\circ$ for $E^{beam}$=5.550~GeV.
Two additional  fits were performed including either (a) $\{\mathbb{H}_{0+},\, \widetilde{\mathbb{H}}_{0+}\}$ to include  genuine twist-3 contributions or (b) $\{\mathbb{H}_{-+},\, \widetilde{\mathbb{H}}_{-+}\}$ to include gluon-transversity GPD contributions. In both of these latter fits $\mathbb{E}_{++}$ and $\widetilde{\mathbb{E}}_{++}$ were set to zero, thus keeping constant the number of free parameters.  The fit to the data is much better ($\chi^2/ndf=210/208$) for both the higher-twist (HT)  or the next-to-leading order (NLO) scenarios than for the LO/LT case.
% Still with $\mathbb{H}_{++}$ and $\tilde{\mathbb{H}}_{++}$, a higher-twist scenario is also displayed in Fig.~\ref{fig:Fit_result} by using $\mathbb{H}_{0+}$ and $\tilde{\mathbb{H}}_{0+}$ to include some genuine twist-3 contributions. Finally a next-to-leading order (NLO) scenario is also tested with $\mathbb{H}_{-+}$ and $\tilde{\mathbb{H}}_{-+}$ to include gluon-transversity GPD contributions, always keeping $\mathbb{H}_{++}$ and $\tilde{\mathbb{H}}_{++}$. For both scenarios, the fit to the data is much better ($\chi^2/Ndf=210/208$) for both scenarios.
This conclusion also holds for the lower $-t$ bins, as summarized in Tab.~\ref{tab:chi}. We observe the crucial role of gluons in the description of the process, either through the quark-gluon correlations involved in higher-twist diagrams, or through next-to-leading effects implying gluon-transversity GPDs. This pioneer analysis including the kinematical power corrections recently calculated for DVCS demonstrate that the leading twist approximation is no longer sufficient to describe the accuracy of these new data.

\begin{table}[tb]
\begin{tabular}{cccc}
%\cline{2-4}
\hline \hline
 Fit Description:  & LO/LT & Higher Twist & NLO \\
 Helicity States:  &++&++/0+&++/$-+$\\
\hline 
$t=-0.18$~GeV$^2$ & $\qquad 314 \qquad$ & $\qquad 251 \qquad$ & $\qquad 256 \qquad$ \\
%\hline
$t=-0.24$~GeV$^2$ & $\qquad 461\qquad$ & $\qquad 246 \qquad$ & $\qquad 245 \qquad$ \\
%\hline
$t=-0.30$~GeV$^2$ & $\qquad 504\qquad$ & $\qquad 211 \qquad$ & $\qquad 210 \qquad$ \\
\hline
\end{tabular}
\caption{Values of $\chi^2$ ($ndf=208$) obtained in the leading-order, leading-twist ($++$); higher-twist ($++$/$0+$); and next-to-leading-order ($++$/$-+$) scenarios. The fit is not performed at the highest value of $-t$ because of the lack of full acceptance in $\phi$, resulting in a large statistical uncertainty.}
\label{tab:chi}
\end{table}

%Although there is no visible difference in the fit result for the total cross section between the twist-3 or the NLO case, the DVCS$^2$ and the interference contributions are strongly different for both polarized and unpolarized cross sections as seen in Fig.~\ref{fig:Fit_result}.

%The DVCS$^2$ and the BH-DVCS interference terms are separated in each of the successful fit scenarios and are presented in Fig.~\ref{fig:sep}. In particular, we note a significant DVCS$^2$ contribution in the higher-twist scenario to the helicity-dependent cross section, assumed to be a purely interference term in DVCS phenomenology up to now. One can see that while both fits yield the same total helicity-dependent and helicity-independent cross sections (Fig.~\ref{fig:Fit_result}), the separated contributions differ significantly for each fit scenario.
%Nonetheless, in either scenario, the interference terms and DVCS$^2$ terms are separated by several times the empirical error band.  Thus our results demonstrate
%the desired sensitivity to both the real and imaginary parts of DVCS amplitude.  We denote this procedure a ``generalized Rosenbluth separation''.

\begin{figure}[htb]
\begin{center}
\includegraphics[width=\linewidth]{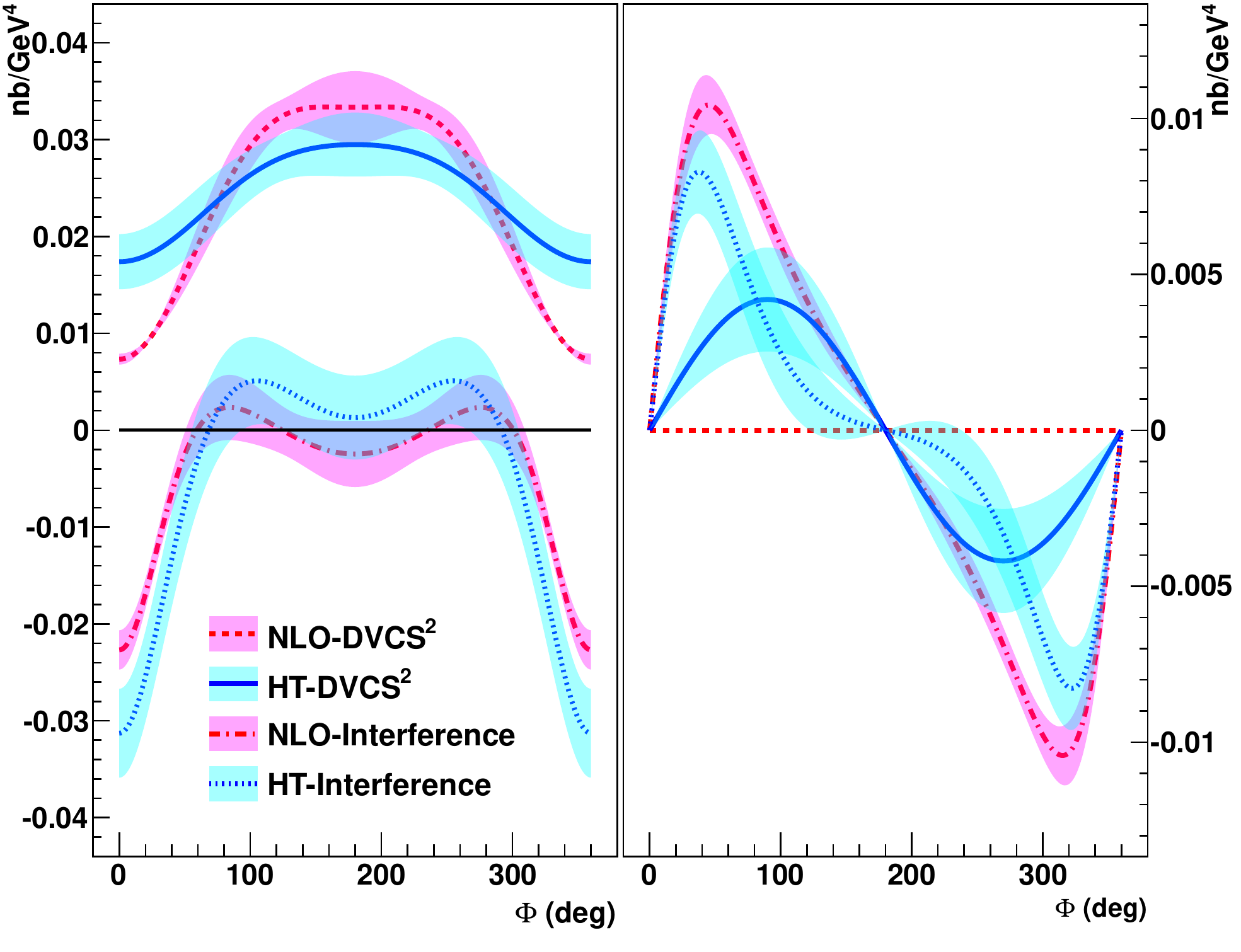}
\caption{\textbf{A generalized Rosenbluth separation.} (Color online) DVCS$^2$ and DVCS-BH interference contributions are shown at $Q^2$=1.75~GeV$^2$, $x_B$=0.36, $t=-0.30$~GeV$^2$ and $E^{beam}$=5.55~GeV for the helicity-independent (left) and helicity-dependent (right) cross sections. Solid and dotted lines represent these contributions for the twist-3 (HT) scenario; dashed and dashed-dotted lines correspond to the NLO scenario. A  DVCS$^2$ contribution appears in the helicity-dependent cross section only if there is a contribution from the longitudinal polarization of the virtual photon (HT scenario).}
\label{fig:sep}
\end{center}
\end{figure}

Within both successful fit scenarios, the DVCS$^2$ and the BH-DVCS interference terms are well separated, as presented in Fig.~\ref{fig:sep}: we denote this procedure a ``generalized Rosenbluth separation''. In particular, we note a significant DVCS$^2$ contribution in the higher-twist scenario to the helicity-dependent cross section, assumed to be a purely interference term in DVCS phenomenology up to now.  In addition, the real part of the BH-DVCS interference (helicity-independent cross section) is extracted
in these kinematics for the first time.

In conclusion, we measured beam helicity-dependent and helicity-independent photon electroproduction cross sections off a proton target for three $Q^2$-values ranging from 1.5 to 2~GeV$^2$ at $x_B$=0.36. Each kinematic setting was measured at two incident beam energies. Using this data set, we demonstrated the sensitivity of high precision DVCS data to twist-3 and/or higher-order contributions through a phenomenological study including for the first time kinematical power corrections. Within either a pure higher-twist or a pure next-to-leading order scenario, both legitimate at our moderate values of $Q^2$, a statistically significant experimental separation of the DVCS$^2$ and DVCS-BH interference terms is achieved. %In the leading-twist and leading-order approach, we performed a fit of the imaginary and real parts of the helicity-conserved CFFs including for the first time kinematical power corrections. This fit is not able to reproduce the data. However, with the same number of degrees of freedom, the agreement between the data and the fit is significantly improved when longitudinal-to-transverse helicity-flip or transverse-to-transverse helicity-flip CFFs are introduced, confirming the sensitivity of high precision DVCS data to twist-3 and/or higher-order contributions. At our moderate values of $Q^2$, both scenarios are legitimate and a separation of the DVCS$^2$ and DVCS-BH interference terms is presented in both of them. A %phenomenological study including the kinematical power corrections over the whole DVCS data might discriminate between the two scenarios. Besides a
Advances in global analyses can include these next-order contributions, rich with information about parton correlations inside the nucleon~\cite{Berthou:2015oaw}. Finally a new program has started at Jefferson Lab to measure deep virtual exclusive scattering with electron beams up to 11 GeV.  For a given $x_B$, the reach in $Q^2$ will span at least a factor of two. This broader reach provides the potential to discriminate between the two scenarios (higher-twist or next-to-leading order), as the cross sections in the two scenarios (for the same GPDs) have different energy and $Q^2$ dependencies at fixed $x_B$.

We thank V.~Braun, M.~Diehl, H.~Moutarde and K.~Kumeri\u{c}ki for valuable discussions about the phenomenological aspects of these results. We acknowledge essential work of the Jefferson Lab accelerator staff and the Hall A technical staff. This work was supported by the Department of Energy (DOE), the National Science Foundation, the French {\em Centre National de la Recherche Scientifique}, the {\em Agence Nationale de la Recherche}, the {\em Commissariat \`a l'\'energie atomique et aux \'energies alternatives} and P2IO Laboratory of Excellence. Jefferson Science Associates, LLC, operates Jefferson Lab for the U.S. DOE under U.S. DOE contract DE-AC05-060R23177.

\bibliography{Compton2014}

\begin{thebibliography}{26}%
\makeatletter
\providecommand \@ifxundefined [1]{%
 \@ifx{#1\undefined}
}%
\providecommand \@ifnum [1]{%
 \ifnum #1\expandafter \@firstoftwo
 \else \expandafter \@secondoftwo
 \fi
}%
\providecommand \@ifx [1]{%
 \ifx #1\expandafter \@firstoftwo
 \else \expandafter \@secondoftwo
 \fi
}%
\providecommand \natexlab [1]{#1}%
\providecommand \enquote  [1]{``#1''}%
\providecommand \bibnamefont  [1]{#1}%
\providecommand \bibfnamefont [1]{#1}%
\providecommand \citenamefont [1]{#1}%
\providecommand \href@noop [0]{\@secondoftwo}%
\providecommand \href [0]{\begingroup \@sanitize@url \@href}%
\providecommand \@href[1]{\@@startlink{#1}\@@href}%
\providecommand \@@href[1]{\endgroup#1\@@endlink}%
\providecommand \@sanitize@url [0]{\catcode `\\12\catcode `\$12\catcode
  `\&12\catcode `\#12\catcode `\^12\catcode `\_12\catcode `\%12\relax}%
\providecommand \@@startlink[1]{}%
\providecommand \@@endlink[0]{}%
\providecommand \url  [0]{\begingroup\@sanitize@url \@url }%
\providecommand \@url [1]{\endgroup\@href {#1}{\urlprefix }}%
\providecommand \urlprefix  [0]{URL }%
\providecommand \Eprint [0]{\href }%
\providecommand \doibase [0]{http://dx.doi.org/}%
\providecommand \selectlanguage [0]{\@gobble}%
\providecommand \bibinfo  [0]{\@secondoftwo}%
\providecommand \bibfield  [0]{\@secondoftwo}%
\providecommand \translation [1]{[#1]}%
\providecommand \BibitemOpen [0]{}%
\providecommand \bibitemStop [0]{}%
\providecommand \bibitemNoStop [0]{.\EOS\space}%
\providecommand \EOS [0]{\spacefactor3000\relax}%
\providecommand \BibitemShut  [1]{\csname bibitem#1\endcsname}%
\let\auto@bib@innerbib\@empty
%</preamble>
\bibitem [{\citenamefont {Mueller}\ \emph {et~al.}(1994)\citenamefont
  {Mueller}, \citenamefont {Robaschik}, \citenamefont {Geyer}, \citenamefont
  {Dittes},\ and\ \citenamefont {Horejsi}}]{Mueller:1998fv}%
  \BibitemOpen
  \bibfield  {author} {\bibinfo {author} {\bibfnamefont {D.}~\bibnamefont
  {Mueller}}, \bibinfo {author} {\bibfnamefont {D.}~\bibnamefont {Robaschik}},
  \bibinfo {author} {\bibfnamefont {B.}~\bibnamefont {Geyer}}, \bibinfo
  {author} {\bibfnamefont {F.~M.}\ \bibnamefont {Dittes}}, \ and\ \bibinfo
  {author} {\bibfnamefont {J.}~\bibnamefont {Horejsi}},\ }\href@noop {}
  {\bibfield  {journal} {\bibinfo  {journal} {Fortschr. Phys.}\ }\textbf
  {\bibinfo {volume} {42}},\ \bibinfo {pages} {101} (\bibinfo {year} {1994})},\
  \Eprint {http://arxiv.org/abs/hep-ph/9812448} {hep-ph/9812448} \BibitemShut
  {NoStop}%
%%CITATION = HEP-PH 9812448;%%
\bibitem [{\citenamefont {Ji}(1997{\natexlab{a}})}]{Ji:1996ek}%
  \BibitemOpen
  \bibfield  {author} {\bibinfo {author} {\bibfnamefont {X.-D.}\ \bibnamefont
  {Ji}},\ }\href@noop {} {\bibfield  {journal} {\bibinfo  {journal} {Phys. Rev.
  Lett.}\ }\textbf {\bibinfo {volume} {78}},\ \bibinfo {pages} {610} (\bibinfo
  {year} {1997}{\natexlab{a}})},\ \Eprint {http://arxiv.org/abs/hep-ph/9603249}
  {hep-ph/9603249} \BibitemShut {NoStop}%
%%CITATION = HEP-PH 9603249;%%
\bibitem [{\citenamefont {Radyushkin}(1997)}]{Radyushkin:1997ki}%
  \BibitemOpen
  \bibfield  {author} {\bibinfo {author} {\bibfnamefont {A.~V.}\ \bibnamefont
  {Radyushkin}},\ }\href@noop {} {\bibfield  {journal} {\bibinfo  {journal}
  {Phys. Rev.}\ }\textbf {\bibinfo {volume} {D56}},\ \bibinfo {pages} {5524}
  (\bibinfo {year} {1997})},\ \Eprint {http://arxiv.org/abs/hep-ph/9704207}
  {hep-ph/9704207} \BibitemShut {NoStop}%
%%CITATION = HEP-PH 9704207;%%
\bibitem [{\citenamefont {Ji}(1997{\natexlab{b}})}]{Ji:1996nm}%
  \BibitemOpen
  \bibfield  {author} {\bibinfo {author} {\bibfnamefont {X.-D.}\ \bibnamefont
  {Ji}},\ }\href {\doibase 10.1103/PhysRevD.55.7114} {\bibfield  {journal}
  {\bibinfo  {journal} {Phys.Rev.}\ }\textbf {\bibinfo {volume} {D55}},\
  \bibinfo {pages} {7114} (\bibinfo {year} {1997}{\natexlab{b}})},\ \Eprint
  {http://arxiv.org/abs/hep-ph/9609381} {arXiv:hep-ph/9609381 [hep-ph]}
  \BibitemShut {NoStop}%
%%CITATION = HEP-PH/9609381;%%
\bibitem [{\citenamefont {Aktas}\ \emph {et~al.}(2005)\citenamefont {Aktas}
  \emph {et~al.}}]{Aktas:2005ty}%
  \BibitemOpen
  \bibfield  {author} {\bibinfo {author} {\bibfnamefont {A.}~\bibnamefont
  {Aktas}} \emph {et~al.} (\bibinfo {collaboration} {H1}),\ }\href@noop {}
  {\bibfield  {journal} {\bibinfo  {journal} {Eur. Phys. J.}\ }\textbf
  {\bibinfo {volume} {C44}},\ \bibinfo {pages} {1} (\bibinfo {year} {2005})},\
  \Eprint {http://arxiv.org/abs/hep-ex/0505061} {hep-ex/0505061} \BibitemShut
  {NoStop}%
%%CITATION = HEP-EX 0505061;%%
\bibitem [{\citenamefont {Adloff}\ \emph {et~al.}(2001)\citenamefont {Adloff}
  \emph {et~al.}}]{Adloff:2001cn}%
  \BibitemOpen
  \bibfield  {author} {\bibinfo {author} {\bibfnamefont {C.}~\bibnamefont
  {Adloff}} \emph {et~al.} (\bibinfo {collaboration} {H1}),\ }\href@noop {}
  {\bibfield  {journal} {\bibinfo  {journal} {Phys. Lett. B}\ }\textbf
  {\bibinfo {volume} {517}},\ \bibinfo {pages} {47} (\bibinfo {year} {2001})},\
  \Eprint {http://arxiv.org/abs/hep-ex/0107005} {hep-ex/0107005} \BibitemShut
  {NoStop}%
%%CITATION = HEP-EX 0107005;%%
\bibitem [{\citenamefont {Chekanov}\ \emph {et~al.}(2003)\citenamefont
  {Chekanov} \emph {et~al.}}]{Chekanov:2003ya}%
  \BibitemOpen
  \bibfield  {author} {\bibinfo {author} {\bibfnamefont {S.}~\bibnamefont
  {Chekanov}} \emph {et~al.} (\bibinfo {collaboration} {ZEUS}),\ }\href@noop {}
  {\bibfield  {journal} {\bibinfo  {journal} {Phys. Lett.}\ }\textbf {\bibinfo
  {volume} {B573}},\ \bibinfo {pages} {46} (\bibinfo {year} {2003})},\ \Eprint
  {http://arxiv.org/abs/hep-ex/0305028} {hep-ex/0305028} \BibitemShut {NoStop}%
%%CITATION = HEP-EX 0305028;%%
\bibitem [{\citenamefont {Airapetian}\ \emph {et~al.}(2008)\citenamefont
  {Airapetian} \emph {et~al.}}]{Airapetian:2008aa}%
  \BibitemOpen
  \bibfield  {author} {\bibinfo {author} {\bibfnamefont {A.}~\bibnamefont
  {Airapetian}} \emph {et~al.} (\bibinfo {collaboration} {HERMES
  Collaboration}),\ }\href {\doibase 10.1088/1126-6708/2008/06/066} {\bibfield
  {journal} {\bibinfo  {journal} {JHEP}\ }\textbf {\bibinfo {volume} {0806}},\
  \bibinfo {pages} {066} (\bibinfo {year} {2008})},\ \Eprint
  {http://arxiv.org/abs/0802.2499} {arXiv:0802.2499 [hep-ex]} \BibitemShut
  {NoStop}%
%%CITATION = ARXIV:0802.2499;%%
\bibitem [{\citenamefont {Fuchey}\ \emph {et~al.}(2011)\citenamefont {Fuchey},
  \citenamefont {Camsonne}, \citenamefont {Munoz~Camacho}, \citenamefont
  {Mazouz}, \citenamefont {Gavalian} \emph {et~al.}}]{Collaboration:2010kna}%
  \BibitemOpen
  \bibfield  {author} {\bibinfo {author} {\bibfnamefont {E.}~\bibnamefont
  {Fuchey}}, \bibinfo {author} {\bibfnamefont {A.}~\bibnamefont {Camsonne}},
  \bibinfo {author} {\bibfnamefont {C.}~\bibnamefont {Munoz~Camacho}}, \bibinfo
  {author} {\bibfnamefont {M.}~\bibnamefont {Mazouz}}, \bibinfo {author}
  {\bibfnamefont {G.}~\bibnamefont {Gavalian}},  \emph {et~al.},\ }\href
  {\doibase 10.1103/PhysRevC.83.025201} {\bibfield  {journal} {\bibinfo
  {journal} {Phys.Rev.}\ }\textbf {\bibinfo {volume} {C83}},\ \bibinfo {pages}
  {025201} (\bibinfo {year} {2011})},\ \Eprint {http://arxiv.org/abs/1003.2938}
  {arXiv:1003.2938 [nucl-ex]} \BibitemShut {NoStop}%
%%CITATION = ARXIV:1003.2938;%%
\bibitem [{\citenamefont {Mazouz}\ \emph {et~al.}(2007)\citenamefont {Mazouz}
  \emph {et~al.}}]{Mazouz:2007aa}%
  \BibitemOpen
  \bibfield  {author} {\bibinfo {author} {\bibfnamefont {M.}~\bibnamefont
  {Mazouz}} \emph {et~al.} (\bibinfo {collaboration} {Jefferson Lab Hall A
  Collaboration}),\ }\href {\doibase 10.1103/PhysRevLett.99.242501} {\bibfield
  {journal} {\bibinfo  {journal} {Phys.Rev.Lett.}\ }\textbf {\bibinfo {volume}
  {99}},\ \bibinfo {pages} {242501} (\bibinfo {year} {2007})},\ \Eprint
  {http://arxiv.org/abs/0709.0450} {arXiv:0709.0450 [nucl-ex]} \BibitemShut
  {NoStop}%
%%CITATION = ARXIV:0709.0450;%%
\bibitem [{\citenamefont {Defurne}\ \emph {et~al.}(2015)\citenamefont {Defurne}
  \emph {et~al.}}]{Defurne:2015kxq}%
  \BibitemOpen
  \bibfield  {author} {\bibinfo {author} {\bibfnamefont {M.}~\bibnamefont
  {Defurne}} \emph {et~al.} (\bibinfo {collaboration} {Jefferson Lab Hall A}),\
  }\href {\doibase 10.1103/PhysRevC.92.055202} {\bibfield  {journal} {\bibinfo
  {journal} {Phys. Rev.}\ }\textbf {\bibinfo {volume} {C92}},\ \bibinfo {pages}
  {055202} (\bibinfo {year} {2015})},\ \Eprint
  {http://arxiv.org/abs/1504.05453} {arXiv:1504.05453 [nucl-ex]} \BibitemShut
  {NoStop}%
%%CITATION = ARXIV:1504.05453;%%
\bibitem [{\citenamefont {Jo}\ \emph {et~al.}(2015)\citenamefont {Jo} \emph
  {et~al.}}]{Jo:2015ema}%
  \BibitemOpen
  \bibfield  {author} {\bibinfo {author} {\bibfnamefont {H.~S.}\ \bibnamefont
  {Jo}} \emph {et~al.} (\bibinfo {collaboration} {CLAS}),\ }\href {\doibase
  10.1103/PhysRevLett.115.212003} {\bibfield  {journal} {\bibinfo  {journal}
  {Phys. Rev. Lett.}\ }\textbf {\bibinfo {volume} {115}},\ \bibinfo {pages}
  {212003} (\bibinfo {year} {2015})},\ \Eprint
  {http://arxiv.org/abs/1504.02009} {arXiv:1504.02009 [hep-ex]} \BibitemShut
  {NoStop}%
%%CITATION = ARXIV:1504.02009;%%
\bibitem [{\citenamefont {Collins}\ \emph {et~al.}(1997)\citenamefont
  {Collins}, \citenamefont {Frankfurt},\ and\ \citenamefont
  {Strikman}}]{Collins:1996fb}%
  \BibitemOpen
  \bibfield  {author} {\bibinfo {author} {\bibfnamefont {J.~C.}\ \bibnamefont
  {Collins}}, \bibinfo {author} {\bibfnamefont {L.}~\bibnamefont {Frankfurt}},
  \ and\ \bibinfo {author} {\bibfnamefont {M.}~\bibnamefont {Strikman}},\
  }\href {\doibase 10.1103/PhysRevD.56.2982} {\bibfield  {journal} {\bibinfo
  {journal} {Phys.Rev.}\ }\textbf {\bibinfo {volume} {D56}},\ \bibinfo {pages}
  {2982} (\bibinfo {year} {1997})},\ \Eprint
  {http://arxiv.org/abs/hep-ph/9611433} {arXiv:hep-ph/9611433 [hep-ph]}
  \BibitemShut {NoStop}%
%%CITATION = HEP-PH/9611433;%%
\bibitem [{\citenamefont {Ji}\ and\ \citenamefont {Osborne}(1998)}]{Ji:1998xh}%
  \BibitemOpen
  \bibfield  {author} {\bibinfo {author} {\bibfnamefont {X.-D.}\ \bibnamefont
  {Ji}}\ and\ \bibinfo {author} {\bibfnamefont {J.}~\bibnamefont {Osborne}},\
  }\href@noop {} {\bibfield  {journal} {\bibinfo  {journal} {Phys. Rev.}\
  }\textbf {\bibinfo {volume} {D58}},\ \bibinfo {pages} {094018} (\bibinfo
  {year} {1998})},\ \Eprint {http://arxiv.org/abs/hep-ph/9801260}
  {hep-ph/9801260} \BibitemShut {NoStop}%
%%CITATION = HEP-PH 9801260;%%
\bibitem [{\citenamefont {Belitsky}\ \emph {et~al.}(2014)\citenamefont
  {Belitsky}, \citenamefont {Mueller},\ and\ \citenamefont
  {Ji}}]{Belitsky:2012ch}%
  \BibitemOpen
  \bibfield  {author} {\bibinfo {author} {\bibfnamefont {A.~V.}\ \bibnamefont
  {Belitsky}}, \bibinfo {author} {\bibfnamefont {D.}~\bibnamefont {Mueller}}, \
  and\ \bibinfo {author} {\bibfnamefont {Y.}~\bibnamefont {Ji}},\ }\href
  {\doibase 10.1016/j.nuclphysb.2013.11.014} {\bibfield  {journal} {\bibinfo
  {journal} {Nucl. Phys.}\ }\textbf {\bibinfo {volume} {B878}},\ \bibinfo
  {pages} {214} (\bibinfo {year} {2014})},\ \Eprint
  {http://arxiv.org/abs/1212.6674} {arXiv:1212.6674 [hep-ph]} \BibitemShut
  {NoStop}%
%%CITATION = ARXIV:1212.6674;%%
\bibitem [{\citenamefont {Belitsky}\ and\ \citenamefont
  {Mueller}(2010)}]{Belitsky:2010jw}%
  \BibitemOpen
  \bibfield  {author} {\bibinfo {author} {\bibfnamefont {A.}~\bibnamefont
  {Belitsky}}\ and\ \bibinfo {author} {\bibfnamefont {D.}~\bibnamefont
  {Mueller}},\ }\href {\doibase 10.1103/PhysRevD.82.074010} {\bibfield
  {journal} {\bibinfo  {journal} {Phys.Rev.}\ }\textbf {\bibinfo {volume}
  {D82}},\ \bibinfo {pages} {074010} (\bibinfo {year} {2010})},\ \Eprint
  {http://arxiv.org/abs/1005.5209} {arXiv:1005.5209 [hep-ph]} \BibitemShut
  {NoStop}%
%%CITATION = ARXIV:1005.5209;%%
\bibitem [{\citenamefont {Bacchetta}\ \emph {et~al.}(2004)\citenamefont
  {Bacchetta}, \citenamefont {D'Alesio}, \citenamefont {Diehl},\ and\
  \citenamefont {Miller}}]{Bacchetta:2004jz}%
  \BibitemOpen
  \bibfield  {author} {\bibinfo {author} {\bibfnamefont {A.}~\bibnamefont
  {Bacchetta}}, \bibinfo {author} {\bibfnamefont {U.}~\bibnamefont {D'Alesio}},
  \bibinfo {author} {\bibfnamefont {M.}~\bibnamefont {Diehl}}, \ and\ \bibinfo
  {author} {\bibfnamefont {C.~A.}\ \bibnamefont {Miller}},\ }\href@noop {}
  {\bibfield  {journal} {\bibinfo  {journal} {Phys. Rev.}\ }\textbf {\bibinfo
  {volume} {D70}},\ \bibinfo {pages} {117504} (\bibinfo {year} {2004})},\
  \Eprint {http://arxiv.org/abs/hep-ph/0410050} {hep-ph/0410050} \BibitemShut
  {NoStop}%
%%CITATION = HEP-PH 0410050;%%
\bibitem [{\citenamefont {Kumericki}\ and\ \citenamefont
  {Mueller}(2010)}]{Kumericki:2009uq}%
  \BibitemOpen
  \bibfield  {author} {\bibinfo {author} {\bibfnamefont {K.}~\bibnamefont
  {Kumericki}}\ and\ \bibinfo {author} {\bibfnamefont {D.}~\bibnamefont
  {Mueller}},\ }\href {\doibase 10.1016/j.nuclphysb.2010.07.015} {\bibfield
  {journal} {\bibinfo  {journal} {Nucl.Phys.}\ }\textbf {\bibinfo {volume}
  {B841}},\ \bibinfo {pages} {1} (\bibinfo {year} {2010})},\ \Eprint
  {http://arxiv.org/abs/0904.0458} {arXiv:0904.0458 [hep-ph]} \BibitemShut
  {NoStop}%
%%CITATION = ARXIV:0904.0458;%%
\bibitem [{\citenamefont {Kumericki}\ \emph {et~al.}(2016)\citenamefont
  {Kumericki}, \citenamefont {Liuti},\ and\ \citenamefont
  {Moutarde}}]{Kumericki:2016ehc}%
  \BibitemOpen
  \bibfield  {author} {\bibinfo {author} {\bibfnamefont {K.}~\bibnamefont
  {Kumericki}}, \bibinfo {author} {\bibfnamefont {S.}~\bibnamefont {Liuti}}, \
  and\ \bibinfo {author} {\bibfnamefont {H.}~\bibnamefont {Moutarde}},\ }\href
  {\doibase 10.1140/epja/i2016-16157-3} {\bibfield  {journal} {\bibinfo
  {journal} {Eur. Phys. J.}\ }\textbf {\bibinfo {volume} {A52}},\ \bibinfo
  {pages} {157} (\bibinfo {year} {2016})},\ \Eprint
  {http://arxiv.org/abs/1602.02763} {arXiv:1602.02763 [hep-ph]} \BibitemShut
  {NoStop}%
%%CITATION = ARXIV:1602.02763;%%
\bibitem [{\citenamefont {Dupre}\ \emph {et~al.}(2017)\citenamefont {Dupre},
  \citenamefont {Guidal},\ and\ \citenamefont {Vanderhaeghen}}]{Dupre:2016mai}%
  \BibitemOpen
  \bibfield  {author} {\bibinfo {author} {\bibfnamefont {R.}~\bibnamefont
  {Dupre}}, \bibinfo {author} {\bibfnamefont {M.}~\bibnamefont {Guidal}}, \
  and\ \bibinfo {author} {\bibfnamefont {M.}~\bibnamefont {Vanderhaeghen}},\
  }\href {\doibase 10.1103/PhysRevD.95.011501} {\bibfield  {journal} {\bibinfo
  {journal} {Phys. Rev.}\ }\textbf {\bibinfo {volume} {D95}},\ \bibinfo {pages}
  {011501} (\bibinfo {year} {2017})},\ \Eprint
  {http://arxiv.org/abs/1606.07821} {arXiv:1606.07821 [hep-ph]} \BibitemShut
  {NoStop}%
%%CITATION = ARXIV:1606.07821;%%
\bibitem [{\citenamefont {Aaron}\ \emph {et~al.}(2009)\citenamefont {Aaron}
  \emph {et~al.}}]{Aaron:2009ac}%
  \BibitemOpen
  \bibfield  {author} {\bibinfo {author} {\bibfnamefont {F.}~\bibnamefont
  {Aaron}} \emph {et~al.} (\bibinfo {collaboration} {H1 Collaboration}),\
  }\href {\doibase 10.1016/j.physletb.2009.10.035} {\bibfield  {journal}
  {\bibinfo  {journal} {Phys.Lett.}\ }\textbf {\bibinfo {volume} {B681}},\
  \bibinfo {pages} {391} (\bibinfo {year} {2009})},\ \Eprint
  {http://arxiv.org/abs/0907.5289} {arXiv:0907.5289 [hep-ex]} \BibitemShut
  {NoStop}%
%%CITATION = ARXIV:0907.5289;%%
\bibitem [{\citenamefont {Alcorn}\ \emph {et~al.}(2004)\citenamefont {Alcorn}
  \emph {et~al.}}]{Alcorn:2004sb}%
  \BibitemOpen
  \bibfield  {author} {\bibinfo {author} {\bibfnamefont {J.}~\bibnamefont
  {Alcorn}} \emph {et~al.},\ }\href@noop {} {\bibfield  {journal} {\bibinfo
  {journal} {Nucl. Instrum. Meth.}\ }\textbf {\bibinfo {volume} {A522}},\
  \bibinfo {pages} {294} (\bibinfo {year} {2004})}\BibitemShut {NoStop}%
\bibitem [{\citenamefont {Vanderhaeghen}\ \emph {et~al.}(2000)\citenamefont
  {Vanderhaeghen}, \citenamefont {Friedrich}, \citenamefont {Lhuillier},
  \citenamefont {Marchand}, \citenamefont {Van~Hoorebeke},\ and\ \citenamefont
  {Van~de Wiele}}]{Vanderhaeghen:2000ws}%
  \BibitemOpen
  \bibfield  {author} {\bibinfo {author} {\bibfnamefont {M.}~\bibnamefont
  {Vanderhaeghen}}, \bibinfo {author} {\bibfnamefont {J.~M.}\ \bibnamefont
  {Friedrich}}, \bibinfo {author} {\bibfnamefont {D.}~\bibnamefont
  {Lhuillier}}, \bibinfo {author} {\bibfnamefont {D.}~\bibnamefont {Marchand}},
  \bibinfo {author} {\bibfnamefont {L.}~\bibnamefont {Van~Hoorebeke}}, \ and\
  \bibinfo {author} {\bibfnamefont {J.}~\bibnamefont {Van~de Wiele}},\
  }\href@noop {} {\bibfield  {journal} {\bibinfo  {journal} {Phys. Rev.}\
  }\textbf {\bibinfo {volume} {C62}},\ \bibinfo {pages} {025501} (\bibinfo
  {year} {2000})},\ \Eprint {http://arxiv.org/abs/hep-ph/0001100}
  {hep-ph/0001100} \BibitemShut {NoStop}%
%%CITATION = HEP-PH 0001100;%%
\bibitem [{\citenamefont {Rvachev}(2001)}]{Rvachev:2001}%
  \BibitemOpen
  \bibfield  {author} {\bibinfo {author} {\bibfnamefont {M.}~\bibnamefont
  {Rvachev}},\ }\href
  {http://hallaweb.jlab.org/publications/Technotes/technote.html} {\emph
  {\bibinfo {title} {Effective Use of Hall A Spectrometers with
  R-Functions}}},\ \bibinfo {type} {Hall A Technical Note}\ \bibinfo {number}
  {Jlab-TN-01-055}\ (\bibinfo  {institution} {Jefferson Lab},\ \bibinfo {year}
  {2001})\BibitemShut {NoStop}%
\bibitem [{\citenamefont {Braun}\ \emph {et~al.}(2014)\citenamefont {Braun},
  \citenamefont {Manashov}, \citenamefont {Mueller},\ and\ \citenamefont
  {Pirnay}}]{Braun:2014sta}%
  \BibitemOpen
  \bibfield  {author} {\bibinfo {author} {\bibfnamefont {V.~M.}\ \bibnamefont
  {Braun}}, \bibinfo {author} {\bibfnamefont {A.~N.}\ \bibnamefont {Manashov}},
  \bibinfo {author} {\bibfnamefont {D.}~\bibnamefont {Mueller}}, \ and\
  \bibinfo {author} {\bibfnamefont {B.~M.}\ \bibnamefont {Pirnay}},\ }\href
  {\doibase 10.1103/PhysRevD.89.074022} {\bibfield  {journal} {\bibinfo
  {journal} {Phys. Rev.}\ }\textbf {\bibinfo {volume} {D89}},\ \bibinfo {pages}
  {074022} (\bibinfo {year} {2014})},\ \Eprint {http://arxiv.org/abs/1401.7621}
  {arXiv:1401.7621 [hep-ph]} \BibitemShut {NoStop}%
%%CITATION = ARXIV:1401.7621;%%
\bibitem [{\citenamefont {Berthou}\ \emph {et~al.}(2015)\citenamefont
  {Berthou}, \citenamefont {Binosi}, \citenamefont {Chouika}, \citenamefont
  {Guidal}, \citenamefont {Mezrag}, \citenamefont {Moutarde}, \citenamefont
  {Sabati\'{e}}, \citenamefont {Sznajder},\ and\ \citenamefont
  {Wagner}}]{Berthou:2015oaw}%
  \BibitemOpen
  \bibfield  {author} {\bibinfo {author} {\bibfnamefont {B.}~\bibnamefont
  {Berthou}}, \bibinfo {author} {\bibfnamefont {D.}~\bibnamefont {Binosi}},
  \bibinfo {author} {\bibfnamefont {N.}~\bibnamefont {Chouika}}, \bibinfo
  {author} {\bibfnamefont {M.}~\bibnamefont {Guidal}}, \bibinfo {author}
  {\bibfnamefont {C.}~\bibnamefont {Mezrag}}, \bibinfo {author} {\bibfnamefont
  {H.}~\bibnamefont {Moutarde}}, \bibinfo {author} {\bibfnamefont
  {F.}~\bibnamefont {Sabati\'{e}}}, \bibinfo {author} {\bibfnamefont
  {P.}~\bibnamefont {Sznajder}}, \ and\ \bibinfo {author} {\bibfnamefont
  {J.}~\bibnamefont {Wagner}},\ }\href@noop {} {\  (\bibinfo {year} {2015})},\
  \Eprint {http://arxiv.org/abs/1512.06174} {arXiv:1512.06174 [hep-ph]}
  \BibitemShut {NoStop}%
%%CITATION = ARXIV:1512.06174;%%
\end{thebibliography}%

\end{document}